\documentclass[11pt]{article}
\usepackage[margin=1in]{geometry}
\usepackage{amsmath,amssymb,graphicx,booktabs,hyperref,natbib}
\usepackage[T1]{fontenc}
\graphicspath{{figures/}}

\newcommand{\coderepo}{\url{https://github.com/hanssmail/cross-market-liquidity-abm}}

\newcommand{\fnone}{\phi_{\varnothing}}

\title{Herding and Liquidity in Order-Book Markets. II.\\
Fundamental Anchoring and the Resilience of Liquidity}
\author{Jan Novotny\thanks{The views and opinions expressed in this paper are solely those of the author and do not necessarily reflect those of any affiliated institution.}\\
Centre for Econometric Analysis, Bayes Business School\\
106 Bunhill Row, London EC1Y 8TZ, United Kingdom\\
\texttt{jan@novotny.one}}
\date{\today}

\begin{document}
\maketitle

\begin{abstract}
An order-book market whose liquidity provision is anchored to a fundamental value carries a
restoring force: the price mean-reverts to value and the book refills after a shock. We show this
restoring force is a robust intrinsic stabiliser and identify it causally--dialling the anchor
down removes the mean-reversion, and a leverage-driven fire-sale then self-sustains. Separately, we
ask whether a stressed market transmits its liquidity stress to a coupled calmer one, and find that
it cannot: across six transmission channels of increasing strength--cross-market herding, arbitrage
flow, and market-maker withdrawal up to a funding-constrained population fire-sale and a leverage
spiral--the receiver's stress is independent of whether its neighbour is stressed, at every anchor
strength. Market-maker withdrawal thins the receiving book but does not ignite it. Our order
parameter throughout is the one-sidedness of the book--liquidity stress rather than a directional
price crash--so a liquidity crisis here means the sustained one-sidedness a failing anchor produces.
A liquidity crisis in this model is a failure of fundamental anchoring, not of market making.
\end{abstract}


\section{Introduction}

When one market seizes up, does the market next to it seize up too? The transmission of liquidity
stress between markets--a crisis that begins in one order book and spreads to another that was,
on its own, perfectly calm--is one of the mechanisms through which local trouble becomes
systemic. A large theoretical and empirical literature treats such contagion as a near-inevitable
consequence of the links between markets: shared arbitrageurs whose wealth shocks propagate across
their positions \citep{kylexiong2001contagion}, common liquidity providers whose funding
constraints bind simultaneously \citep{brunnermeierpedersen2009funding}, and forced sellers whose
deleveraging drives correlated price impact through several books at once
\citep{contwagalath2013running}. The natural question for a builder of market models is not whether
these links exist--they plainly do--but what, if anything, holds a market together in spite of
them.

This paper answers that question for a minimal but fully microstructural vehicle, and the answer
separates into two findings that the contagion literature tends to conflate. The first is a
positive, causally confirmed result about a market's \emph{intrinsic} stability. The structural
feature that governs it is \emph{fundamental-value anchoring of liquidity provision}: the resting
liquidity in the book is posted around a slowly varying fundamental value rather than around the
current price. Anchoring in this sense is a restoring force. When a shock pushes the price away from
the fundamental, fresh two-sided depth reappears at the fundamental and pulls the price back, so the
price mean-reverts to value and the book refills after the shock. Weaken the anchor and this
restoring force disappears: the market loses its mean-reversion, and under a leverage-driven
fire-sale a shock no longer decays but self-sustains. The anchor is therefore the intrinsic
stabiliser, and we confirm its causal role by dialling it away.

The second finding is a clean null about transmission \emph{between} markets. We couple two such
markets, drive one into a herding-driven liquidity-stress regime, and ask whether the stress crosses
into the calm, sub-critical neighbour through a hierarchy of progressively stronger and more
realistic channels: a cross-market herding signal, an arbitrage order-flow coupling, a single
risk-linked market maker, a funding-constrained market-maker population that synchronously withdraws
its quotes, a demand-side forced fire-sale, and finally a population of leveraged position-holders
whose margin calls set off a self-reinforcing selling spiral. Each channel is stronger than the
last, and several do something genuinely crisis-like--a whole population of market makers pulls its
depth and exits the calm market because a different market is stressed. Yet the receiver's stress
never depends on whether the source is genuinely stressed: no channel transmits, at every anchor
strength, including zero anchor. Cross-market contagion is absent here not because the anchor blocks
it, but because none of these channels carries the source's stress \emph{state} into the receiver.

The two findings are separable, and it is important not to fuse them. Weakening the anchor makes the
receiver \emph{intrinsically} fragile--unstable on its own, even when its neighbour is completely
calm--but it still does not make the receiver catch a stressed neighbour's stress: at every anchor
strength the receiver's instability is statistically identical whether the source is stressed or
calm. The anchor governs intrinsic stability; the absence of contagion is a separate fact that holds
regardless of the anchor. This separability also limits the scope of the positive claim. To obtain a
receiver that could in principle catch a neighbour's stress one must weaken the anchor, and the
weakened receiver is then intrinsically unstable under a calm neighbour; the vehicle therefore
cannot exhibit a market that is stable alone yet infected by its neighbour, and the channel
hierarchy delivers transmission \emph{flow}, never a genuine state-carrying contagion. That
fundamentalist value-anchoring stabilises markets is long established \citep{luxmarchesi1999scaling,
chiarellaiori2002simulation, delong1990noise}; our contribution is the narrower, causal claim: we
isolate the fundamental-value reference of \emph{resting liquidity} as the restoring force by
dialling it away, using a one-sided-book depletion order parameter in a coupled two-market vehicle,
and show that it--not market-maker liquidity withdrawal--is what a liquidity crisis requires to
fail. Alongside it we report a cleanly demonstrated null: across a battery of transmission channels,
and at every anchor strength, no cross-market contagion appears.

Throughout, our order parameter is the one-sidedness of the book--liquidity stress--rather than a
directional price crash; a ``liquidity crisis'' in this paper means the sustained one-sidedness a
failing anchor produces, not a one-directional collapse. The present study is the two-market
companion to a single-market result on the same order-book vehicle, in which correlated momentum
herding drives a null-verified liquidity-stress crossover \citep{novotny2026darkcorners}.

\section{Related work}

Our subject sits at the intersection of four literatures, and we position the stabilising-anchor
finding against each in turn.

\paragraph{Cross-market contagion mechanisms.}
The mechanisms we stress-test are the standard ones. \citet{kylexiong2001contagion} model contagion
as a wealth effect: convergence traders holding positions in two markets are forced to delever in
one when they lose in the other, jointly raising volatility and correlation and lowering liquidity
in both. \citet{brunnermeierpedersen2009funding} tie market liquidity to the funding liquidity of
the agents who provide it, so that a funding shock makes providers withdraw and liquidity dries up
across the markets they serve; \citet{brunnermeierpedersen2005predatory} add the predatory-trading
amplifier. \citet{contwagalath2013running} formalise fire-sale contagion, in which forced
deleveraging generates correlated price impact and endogenous cross-asset correlation. These are
reduced-form or continuous-price models. We reproduce each mechanism as an explicit agent in a
discrete order book--a shared arbitrageur, a risk-linked and then funding-constrained market
maker, a forced fire-sale, a leveraged-holder margin spiral--and ask what an anchored receiving
book does when they act on it. The answer, and the reason it differs from these models, is that the
anchor supplies a restoring force those models do not contain.

\paragraph{Limits of arbitrage.}
Treating the arbitrageur as a capital- and aggressiveness-limited agent rather than a frictionless
price-equaliser follows \citet{shleifer1997limits} and \citet{gromb2002equilibrium}, whose
two-segmented-markets setup is the direct ancestor of an arbitrageur that trades both books to close
a price gap. Classic statistical-arbitrage and pairs-trading practice \citep{gatev2006pairstrading,
avellanedalee2010statarb} motivates the gap-closing rule. Our arbitrageur is deliberately of the
same order of magnitude as native flow, which is what exposes the difference between a genuine
transmission of stress and a mechanical consumption of liquidity.

\paragraph{Market making and the value of anchoring.}
Our liquidity providers use the inventory-controlled quoting of \citet{hostoll1981optimal} and
\citet{avellanedastoikov2008hft}, quoting around a reservation price skewed by inventory; the
adverse-selection and flow-toxicity concerns of \citet{glostenmilgrom1985bidask} and
\citet{easley2012flowtoxicity} motivate the withdrawal response we build in. What these frameworks
leave implicit, and what we make central, is the \emph{reference} around which quotes are posted.
When that reference is a slowly varying fundamental rather than the current price, the provider
population supplies a mean-reverting restoring force. That fundamentalists exert such a stabilising
pull while trend-followers destabilise is long established in the fundamentalist--chartist tradition
\citep{luxmarchesi1999scaling, chiarellaiori2002simulation}, with \citet{chiarellaiori2002simulation}
placing value-referenced orders inside a limit order book, and \citet{delong1990noise} showing that
noise-trader risk bounds the restoring force and makes a weakened anchor meaningful. Our
contribution is not the general claim that fundamentals stabilise, but the causal isolation of the
fundamental-value reference of \emph{resting liquidity}: dialling that anchor away under a controlled
sweep, with a one-sided-book depletion order parameter, in a coupled two-market vehicle.

\paragraph{Agent-based and order-book contagion models.}
The closest computational precedents study arbitrage-mediated coupling directly.
\citet{gaoladley2022statistical} sweep an arbitrage-connectivity density and report a non-monotone
threshold, with low connectivity stabilising and high connectivity destabilising; theirs is a
reduced-form network of linked prices rather than a limit-order book with a bid/ask-depletion order
parameter, and the control parameter is arbitrageur density under an exogenous extreme-event shock
rather than an endogenously stressed source. \citet{ciacci2020triangular} build a genuine
agent-based model with market makers and a triangular arbitrageur across three currency-pair books,
establishing that a cross-book arbitrageur agent is a well-posed construct, but their object is
steady-state cross-rate correlation, not liquidity-stress transmission.
\citet{paulin2018flashcrash} study contagion through portfolio and leverage-network overlap between
separate fund agents rather than direct trading in both books. Central-bank and policy modelling of
cross-market liquidity and dealer behaviour \citep{bookstaber2015crisisliquidity,
karvik2018deeds, borsos2025abmcentralbanks, vytelingum2025liquidityrisk} shares our systemic
motivation at a more aggregated scale. On the empirical side, cross-market liquidity commonality and
its dependence on dealer funding \citep{aramonte2020crossmarket, garrison2019crossasset,
bruche2026dealerfunding, gao2024flashcrash} is exactly the phenomenon our stabiliser speaks to.
Methodologically we follow the phase-diagram-first, null-controlled style of recent order-book
statistical-physics work \citep{bouchaud2024navigating, fosset2020endogenous,
benzaquen2017crossimpact, filimonov2012reflexivity, cespa2014illiquidity, grossmanmiller1988liquidity},
in which a liquidity-stress order parameter is mapped against a control parameter and every apparent
effect is checked against a matched null. To our knowledge, the combination of a discrete coupled
order-book model, a hierarchy of explicit contagion channels, a one-sided-book liquidity-stress
order parameter, and a causal dial on fundamental anchoring has not been reported.

\section{Model}
\label{sec:model}

\subsection{The single-market order book and its restoring force}

Each market is a continuous double auction on a discrete price grid. Time advances one event at a
time; at each event an agent submits a limit or market order, and resting limit orders form a
two-sided book with a best bid and a best ask whenever both sides are populated. Two agent classes
generate the native flow.

\emph{Anchored liquidity providers.} A population of zero-intelligence traders posts limit orders on
both sides of the book around a reference price. Their reference is the \emph{fundamental value}
$f_t$, a slowly varying random walk common to all of them. Because every provider quotes around the
same $f_t$, the book is continually replenished with two-sided depth centred on the fundamental.
This is the restoring force at the heart of the paper: whenever trading pushes the transaction price
$p_t$ away from $f_t$, fresh bids and asks reappear at $f_t$ and draw the price back, so
$p_t$ mean-reverts to $f_t$. The replenishing depth is posted at a fixed distance from $f_t$
regardless of how far $p_t$ has strayed; we characterise the resulting mean reversion by its effect
on the order parameter below rather than by a displacement-dependent rate, and refer to it simply as
a restoring force.

\emph{Momentum herders.} A fraction of the flow comes from chartist agents whose order sign follows
the recent price momentum. Two parameters govern them: an intensity $\phi \in [0,1]$, the propensity
to act on the momentum signal, and a coupling $\kappa$ that correlates the herders with one another.
When $\phi$ and $\kappa$ are large the herders act in concert: a run of upticks provokes correlated
buying, a run of downticks correlated selling. Above a smooth onset boundary $\phi^{*}(\kappa)$ this
correlated one-directional pressure repeatedly exhausts one side of the book faster than the
providers refill it, and the book spends an appreciable fraction of time \emph{one-sided}--with
no best bid or no best ask, and hence no defined mid-price.

We measure liquidity stress by the order parameter
\begin{equation}
  \fnone \;=\; \text{long-run fraction of events at which the book is one-sided (no defined mid)}
  \,.
  \label{eq:fnone}
\end{equation}
In a calm, sub-critical market $\fnone \approx 0$; in the herding stress regime one side of the
book is repeatedly empty and $\fnone$ rises to a plateau of order $0.34$. The single-market
crossover in $\fnone$ across $\phi^{*}(\kappa)$, together with the fact that a sign-scrambled null
returns $\fnone = 0$ exactly, is established for this vehicle in the companion single-market study
\citep{novotny2026darkcorners} and is not re-derived here; we take the stressed corner
$(\phi=0.9,\ \kappa=1.0)$ as the
liquidity-stress source and hold receiving markets at sub-critical $\phi$.

\subsection{The anchor dial}
\label{sec:anchordial}

To interrogate the restoring force we replace the providers' fixed fundamental reference by a
convex blend of the fundamental and the current price. For a market with anchor strength
$a \in [0,1]$ the provider reference at event $t$ is
\begin{equation}
  r_t \;=\; a\,f_t \;+\; (1-a)\,p_t \,,
  \label{eq:anchor}
\end{equation}
where $p_t$ is the live mid-price (or, if the book is one-sided, the last trade price, falling back
to $f_t$). At $a=1$ the providers quote around the fundamental and the full restoring force is
present; this is the baseline market of the previous subsection. At $a=0$ they quote around the
current price, so a dislocation is met by fresh depth \emph{at the dislocated level} and there is no
force pulling the price back--the price then random-walks. The dial \eqref{eq:anchor} thus turns
the restoring force continuously on and off while changing nothing else about the model, and at
$a=1$ it reproduces the baseline exactly.

\subsection{Coupling two markets}
\label{sec:coupling}

We couple a source market A and a receiver market B on a common event clock: each tick advances both
books by one event, each drawn from its own independent random-number stream. Market A is held at
the stressed corner and market B is held sub-critical and, unless the anchor dial is engaged, fully
anchored ($a=1$). The coupling between the two books is a single channel at a time, chosen from a
hierarchy of increasing strength. Each channel carries one intensity parameter, and at zero
intensity the two markets are independent, which we verify as a foundation gate in every case. The
channels are as follows.

\emph{(i) Cross-market herding signal.} B's herders react not only to B's own momentum but,
with weight $c$, to A's momentum: the effective signal a B-herder acts on is
$s^{B}_t + c\,s^{A}_t$, where $s^{X}_t$ is market $X$'s momentum signal. Only a scalar signal
crosses; no order flow from A ever touches B's book.

\emph{(ii) Arbitrage order flow.} A cross-market arbitrageur observes the price gap
$\mathrm{gap}_t = p^{A}_t - p^{B}_t$ and, with probability proportional to $g\,|\mathrm{gap}_t|$,
sends a unit market order to each book--selling the richer, buying the cheaper. Here order flow,
not a signal, crosses the two books.

\emph{(iii) Single risk-linked market maker.} One provider quotes both books with
inventory-skewed two-sided quotes. It carries a shared-risk weight $\lambda\in[0,1]$ that blends its
per-book risk with the pooled risk of both books,
$R^{X}_{\mathrm{eff}} = (1-\lambda)R^{X}_{\mathrm{local}} + \lambda\bigl(R^{A}_{\mathrm{local}}+
R^{B}_{\mathrm{local}}\bigr)$, and widens its spread and cuts its quoted size in proportion to
$R^{X}_{\mathrm{eff}}$. At $\lambda=1$ a stressed A raises the risk the maker perceives in the calm
B, so it withdraws liquidity from B because A is stressed.

\emph{(iv) Funding-constrained market-maker population.} A population of $N$ such makers quotes both
books, now coupled through a funding constraint in the manner of \citet{brunnermeierpedersen2009funding}.
Each maker marks its equity to market, computes a per-book margin that rises with the book's realised
volatility, and forms a funding utilisation $u_i$ equal to total margin over equity. Rising A-volatility
raises $u_i$ for \emph{every} maker at once, so the population's withdrawal is synchronous; when $u_i$
crosses an exit threshold the maker withdraws from both books to an unwind-only state, re-entering only
after its utilisation falls and its equity recovers. Total baseline quoting capacity is held fixed as
$N$ varies.

\emph{(v) Demand-side forced fire-sale.} A maker that crosses its funding-exit threshold now also
force-sells a fraction $L\in[0,1]$ of its B inventory each tick through marketable orders, consuming
the thinned book from the demand side \citep{contwagalath2013running}. Because all makers see the
same stressed A, the fire-sale is synchronous.

\emph{(vi) Leveraged-holder margin spiral.} A population of $H$ directional long-B position-holders
is financed at leverage $\Lambda$. Each holder marks its position to B's price $m$, holds equity
$\text{pos}\cdot m - \text{debt}$, and is forced to market-sell part of its position when its margin
ratio falls below a maintenance level. Selling into a thin, falling book realises a loss, which cuts
equity, which forces more holders to breach and sell: a self-reinforcing fire-sale spiral in the
sense of \citet{contwagalath2013running}. The cross-market priming is channel (iv): A-stress makes
the maker population withdraw from B, thinning the book, raising price impact per share, and lowering
the spiral's ignition threshold.

Channels (iii)--(vi) share the same inventory-controlled quoting mechanics
\citep{hostoll1981optimal, avellanedastoikov2008hft}. Each channel is a strict superset of the
transmission capacity of the one before it, so the hierarchy is a monotone strengthening of the
would-be contagion mechanism.

\section{Experimental setup}
\label{sec:setup}

\subsection{Protocol}

Every experiment couples a stressed source market A--held at the herding corner
$(\phi_A=0.9,\ \kappa=1.0)$, where its own $\fnone^{A}\approx 0.34$--to a calm receiver market B
held sub-critical (typically $\phi_B=0.35$), and asks whether the coupling raises the receiver's
liquidity-stress order parameter $\fnone^{B}$. Signal-only runs use long records of $1.5\times 10^{5}$
events with the first $2\times 10^{4}$ discarded as burn-in; the market-maker, fire-sale, leverage,
and anchor experiments use matched shorter records chosen so that the population computations remain
tractable. The momentum signal is a windowed price change over $200$ events. In the decisive
control the stressed and calm conditions deliberately \emph{share} the same seed set (seeds
$1000$--$1019$), so that each stressed run is paired with a calm run driven by identical native
randomness; this pairing supports a stronger paired per-seed test of the no-contagion null (reported
below) in addition to the unpaired comparison. All reported quantities are means over independent
seeds with the standard error of the mean. The simulation is event-driven, so there is no integrator
step to converge; every run is checked for non-finite values and none occur.

\subsection{Foundation gates}

Before any coupling is interpreted we require two invariants. First, at zero coupling intensity the
two-market system must reduce exactly to two independent single markets, each reproducing the
single-market crossover: at the corner $\fnone^{A}\approx 0.338$ and $\fnone^{B}\approx 0.343$, with
a sign-scrambled null of exactly zero in both. Second, each provider channel must show the correct
sign: a liquidity-\emph{taking} arbitrageur must raise $\fnone$ by draining the book, while a
liquidity-\emph{providing} maker must lower it by adding depth. Both invariants are verified for
every channel; the market-maker population, for example, drives $\fnone^{B}$ from $0.035$ to $0.000$
purely by adding depth. The anchor experiments carry a further gate: at $a=1$ the anchored run must
reproduce the baseline market bit for bit, which we confirm across stressed and calm conditions.

\subsection{The decisive stressed-versus-matched control}

The central methodological hazard is that a coupling agent which \emph{itself} trades in B can raise
$\fnone^{B}$ mechanically--by consuming B's liquidity with its own orders--without transmitting
any of A's stress \emph{state}. We separate the two with a matched control repeated for every
order-flow channel. We measure the coupling's realised order flow into B under a genuinely stressed
A, then reproduce the \emph{same} average flow under a calm A--either by holding A unstressed at
matched intensity, or by injecting an exogenous flow of the measured rate and side while A is calm.
Genuine contagion requires the stressed run to exceed the matched control beyond sampling error; a
mechanical drain gives equality. We report the increment (stressed minus matched control) and its
$z$-score. Because the stressed and matched-calm conditions share seeds, we also run the comparison
as a paired per-seed test: at anchor strengths $a=0.25$, $0.10$, $0.00$ it gives $t\approx-0.47$,
$-0.61$, $-0.53$, confirming the no-contagion null under the stronger paired design. Alongside it we
use an open-loop control (the receiver driven by a frozen recording of a
stressed A, with no feedback path) to test whether any live bidirectional feedback loop is needed,
and, for the leverage spiral, a spiral-isolating control (the full stack minus the withdrawal
priming) that removes the mechanical drain entirely.

\subsection{The anchor sweep}

The causal experiment sweeps the receiver's anchor strength $a$ over $\{1.0, 0.75, 0.5, 0.25, 0.1,
0.0\}$, crossed with holder leverage $\Lambda \in \{4,6,8\}$, under the full coupling stack
(funding-constrained population withdrawal plus leveraged holders plus the margin spiral). At each
anchor strength we run the decisive stressed-versus-calm control at twenty seeds to ask the two
distinct questions the dial separates: whether the receiver becomes unstable \emph{on its own} as
$a\to 0$ (intrinsic fragility), and whether, at any $a$, a stressed A drives more instability than a
calm A (cross-market contagion). We also apply a windowed exogenous sell shock to the receiver, at
each $a$, to measure directly whether a dislocation decays (anchored) or persists (unanchored).

\section{Results}
\label{sec:results}

\subsection{A stabilising restoring force}
\label{sec:restoring}

The anchor supplies a restoring force that keeps the book two-sided. The cleanest demonstration is a
direct shock. We take a single anchored market ($a=1$), hold it otherwise calm, and hit it with a
windowed exogenous burst of sell market orders. During the burst the book goes one-sided and the
order parameter climbs to $\fnone \approx 0.70$--$0.88$; the mechanism is genuinely being driven
into stress. The moment the burst relents the book recovers: the price reverts to the fundamental
\emph{exactly} (the post-shock mid returns to $1000.0$, its fundamental level), and $\fnone$ falls
back to $\approx 0.04$. Fresh two-sided depth reappears at the fundamental faster than the shock can
empty the book, so a dislocation cannot persist. This is the restoring force in action, and
it is the property that every coupling experiment below tests against.

The same restoring force keeps a calm, sub-critical market at $\fnone \approx 0$ even when a
liquidity-providing agent is present: a market-maker population reduces the receiver's stress from
$\fnone^{B} = 0.035$ to $0.000$ by adding anchored depth, and denser provision makes the market more
robust, not less. Any coupling that raised such a market's stress would have to empty the book
faster than the anchor refills it--a demanding bar for the receiver's intrinsic stability. Whether
the transmission channels clear it, and whether the receiver's stress depends on the source at all,
is what the stress test now examines.

\subsection{The stress test: a hierarchy of transmission channels, all absorbed}
\label{sec:stresstest}

We couple two anchored markets, drive A to the stressed corner, and attempt to transmit its stress
into a calm sub-critical B through the six channels of Section~\ref{sec:coupling}, each stronger than
the last. Table~\ref{tab:channels} collects the outcome. Two things move across the hierarchy. The
\emph{withdrawal / transmission mechanism} strengthens monotonically: from a copied signal, to a
liquidity-draining arbitrageur, to a single risk-linked maker that pulls depth, to a whole
funding-constrained population that synchronously exits, to a forced fire-sale, to a self-reinforcing
leverage spiral. The \emph{receiving-market order parameter} $\fnone^{B}$ does not follow: at every
step the anchored receiver absorbs the stress, and the decisive stressed-versus-matched control
never shows a stressed A transmitting more stress than a flow-matched calm A.

\begin{table}[t]
\centering
\caption{The stress-test hierarchy. Each channel is a stronger would-be contagion mechanism than the
one above it. The transmission mechanism strengthens monotonically, yet the anchored receiver's
liquidity-stress order parameter $\fnone^{B}$ stays negligible and A-stress-independent throughout.
The decisive column reports the stressed-minus-matched-control increment and its $z$-score.}
\label{tab:channels}
\small
\begin{tabular}{@{}p{4.4cm}p{4.4cm}p{5.2cm}@{}}
\toprule
Channel & Transmission mechanism & Receiver stress $\fnone^{B}$ \\
\midrule
Cross-market herding signal &
external momentum blend (no order flow) &
stressed $\approx$ amplitude-matched decorrelated; $|z|<1$ at 3 of 4 cells, added push $\le 0.005$
above baseline \\
Arbitrage order flow &
mechanical liquidity drain &
stressed $A \le$ flow-matched unstressed $A$ (mechanical-drain-consistent) \\
Single risk-linked market maker &
real withdrawal: depth $-64\%$, spread $+136\%$ in calm B &
marginal, not significant ($z\approx 1.7$, $p\approx 0.08$), sign-inconsistent \\
Funding-constrained maker population &
real, large, synchronous exit ($\sim 10\to 6.5$ active makers) &
null, increment $0.00016$, $z=1.03$ \\
$+$ demand-side forced fire-sale &
real fire-sale, negligible volume &
null, increment $0.00021$, $z=1.00$ \\
$+$ leveraged-holder margin spiral &
spiral fires under a direct shock, never ignited by A-stress &
null, spiral-isolating increment $0.0004$, $z=0.03$ \\
\bottomrule
\end{tabular}
\end{table}

\paragraph{Signal and order-flow channels.}
When only a momentum signal crosses, a receiver held near its own onset is indifferent to whether
the coupled market is genuinely stressed or a decorrelated signal of equal amplitude: at three of
four amplitude-matched cells the stressed and decorrelated conditions are statistically
indistinguishable ($|z|<1$), and adding a full stressed-A push lifts $\fnone^{B}$ by at most
$0.005$ above its uncoupled baseline, against a crossover scale of $0.34$. An external signal cannot
instantiate the receiver's own reflexive loop. When order flow crosses through an arbitrageur, the
receiver's stress rises with intensity--but so does the source's, in lockstep, the signature of a
mechanical drain rather than selective transmission. The decisive control confirms it: at matched
flow a genuinely stressed A produces no more receiver stress than an unstressed A (if anything
slightly less), and an open-loop recording of a stressed A reproduces the effect with no feedback
loop at all.

\begin{figure}[t]
\centering
\includegraphics[width=0.78\textwidth]{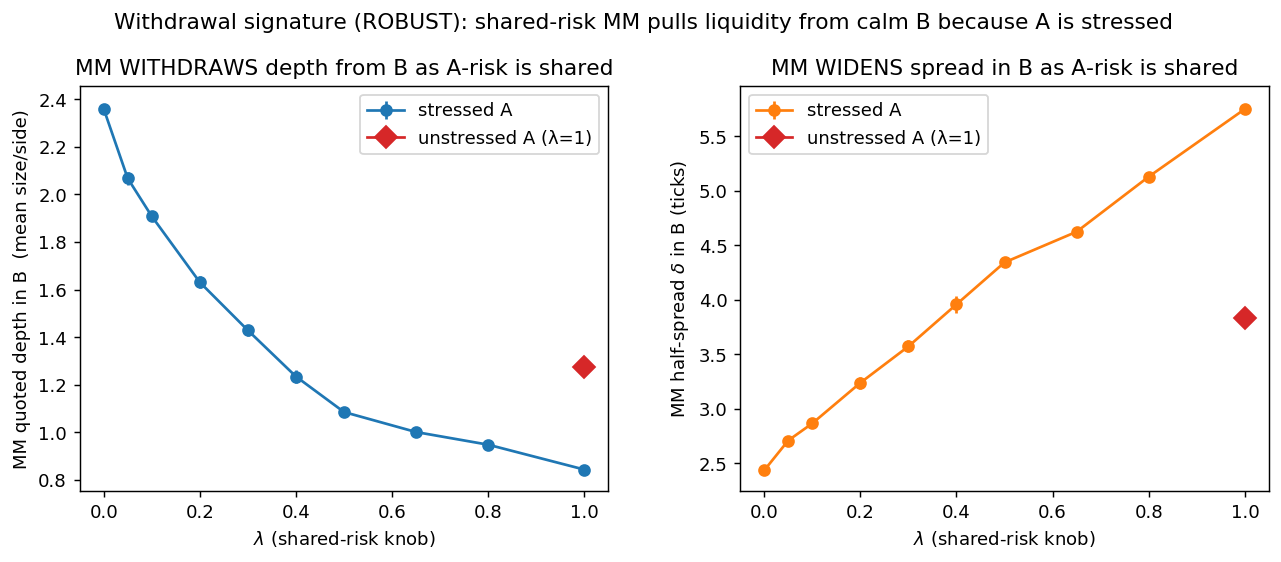}
\caption{The withdrawal mechanism is real. As the shared-risk weight $\lambda$ rises, the market
maker pulls quoted depth (left) and widens its half-spread (right) in the calm receiver B, and does
so more strongly when the source market A is genuinely stressed than when A is calm. The maker
withdraws liquidity from a calm market because a different market is stressed--yet the anchored
receiver's realised stress does not follow.}
\label{fig:withdrawal}
\end{figure}

\begin{figure}[t]
\centering
\includegraphics[width=0.92\textwidth]{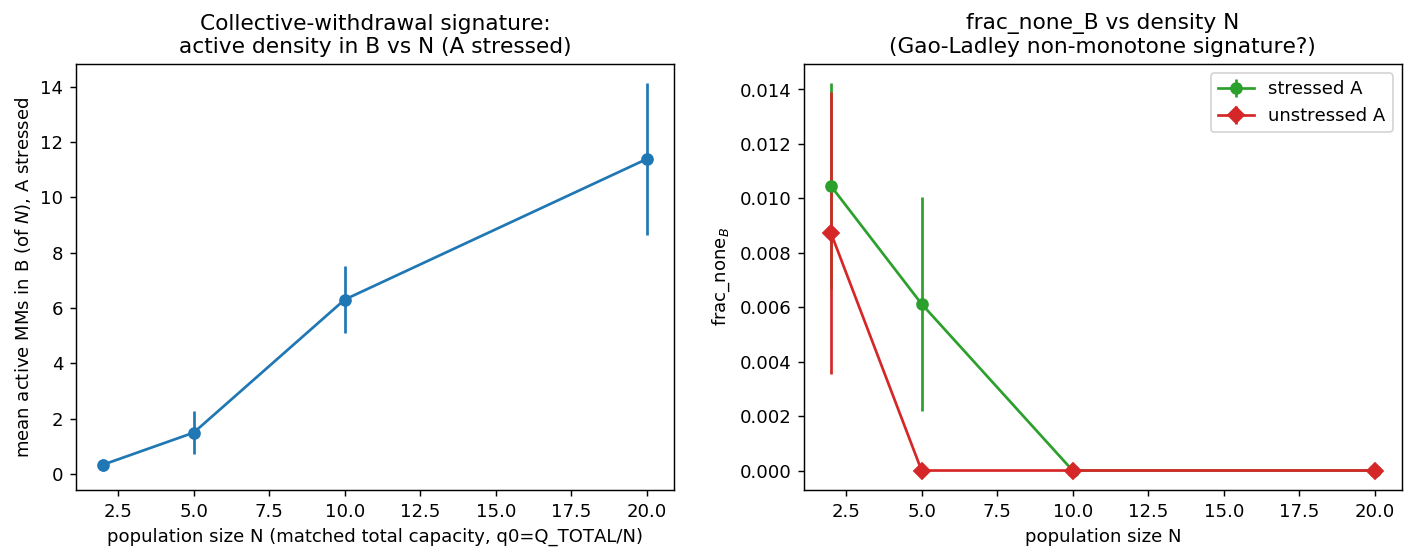}
\caption{A funding-constrained maker population synchronously withdraws, but the receiver stays calm.
Left: under a stressed A the active maker density in the calm receiver B collapses (about $10\to 6.5$
makers, near one in some seeds), entirely absent when A is calm--the strongest, cleanest
A-stress-driven withdrawal in the hierarchy. Right: the receiver's order parameter $\fnone^{B}$
falls monotonically towards zero as maker density rises; more providers make the receiver more
robust, and there is no destabilising density threshold.}
\label{fig:population}
\end{figure}

\paragraph{Withdrawal channels.}
The provider channels do something economically real. A single risk-linked maker, as its shared-risk
weight $\lambda$ is turned up, withdraws liquidity from the calm receiver \emph{because the other
market is stressed}: its quoted depth in B falls by $64\%$ and its half-spread widens by $136\%$, and
it withdraws more when A is genuinely stressed than when A is calm (Figure~\ref{fig:withdrawal}). A
whole funding-constrained population does more still: under a stressed A the population synchronously
exits the calm receiver, its active membership collapsing from $10$ to about $6.5$ makers and, in
some seeds, to near one (Figure~\ref{fig:population}, left). This is a large, clean, unambiguously
A-stress-driven withdrawal. Yet the receiver's realised order parameter does not follow: the
stressed-minus-calm increment in $\fnone^{B}$ is $0.00016$ ($z=1.03$, not significant), carried
entirely by a single rare-excursion seed, and a dynamically exiting population produces the same
receiver stress as a capacity-matched static one. Denser provision makes the receiver \emph{less}
stressed, not more (Figure~\ref{fig:population}, right): there is no destabilising density threshold.
The withdrawal transmits; the receiver's crisis does not follow.

\paragraph{Demand-side and leverage channels.}
Completing the mechanism with a demand-side amplifier does not change the verdict. A forced fire-sale
by the deleveraging makers genuinely fires--gated correctly on A-stress--but injects a
negligible volume, because market makers hold bounded inventory; the receiver does not tip at any
fire-sale aggressiveness, and the stressed run matches a flow-matched drain ($z=1.00$). Adding a
population of leveraged directional holders, whose margin calls drive a self-reinforcing fire-sale
spiral, supplies the one institutional ingredient the model was missing. The spiral is built
correctly: it is calm-stable, and it genuinely fires under a direct exogenous shock, spiking
$\fnone^{B}$ to $\approx 0.88$ during the shock. But under the actual contagion protocol--A
stressed, no exogenous shock--the cross-market priming is too weak to breach even one holder, and
the spiral-isolating control (the full stack minus the withdrawal priming) gives an increment of
$0.0004$ ($z=0.03$): the leverage spiral adds nothing. As Section~\ref{sec:restoring} showed, the
reason is the restoring force. The anchored price reverts to the fundamental, so a forced seller can
always exit near par, leverage losses are bounded, and stress remains a transient one-sidedness of
the book rather than a sustained collapse for a runaway to feed on.

\subsection{Causal confirmation: weakening the anchor}
\label{sec:causal}

\begin{figure}[t]
\centering
\includegraphics[width=0.82\textwidth]{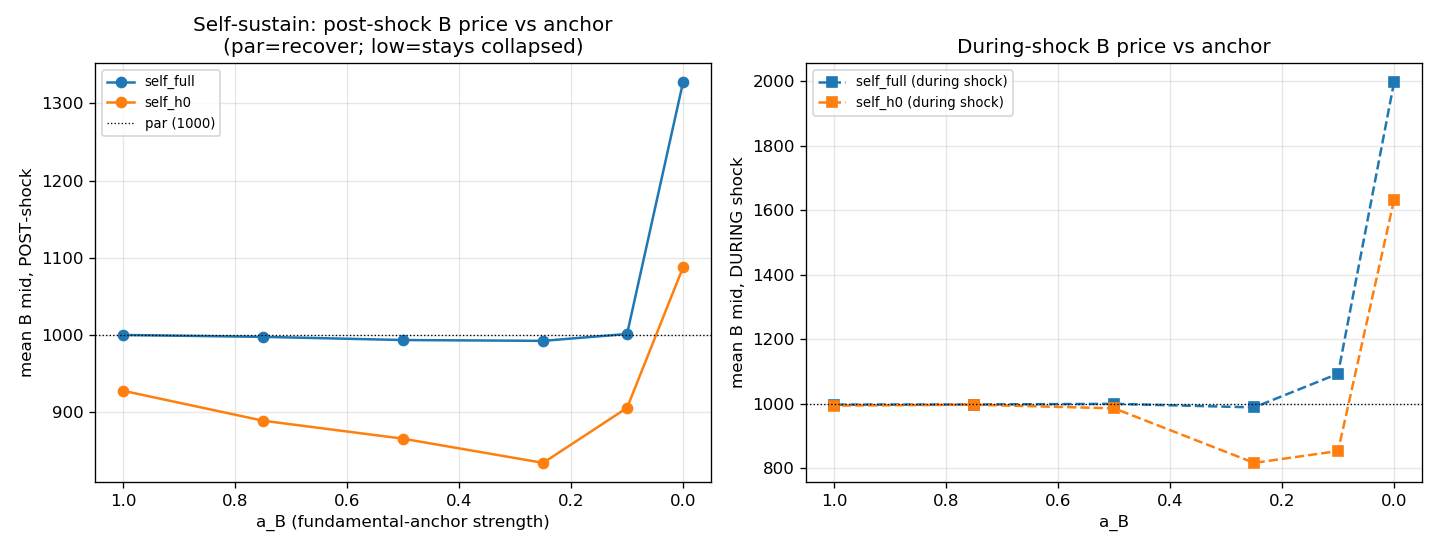}
\caption{The anchor is the restoring force. After a windowed exogenous sell shock, a strongly
anchored receiver ($a=1$) reverts to the fundamental and its one-sidedness collapses back to
$\approx 0.04$; as the anchor is weakened ($a\to 0$) the post-shock one-sidedness rises and persists,
because the price random-walks with no force pulling it back. The dislocation decays when anchored
and self-sustains when unanchored.}
\label{fig:selfsustain}
\end{figure}

\begin{figure}[t]
\centering
\includegraphics[width=0.82\textwidth]{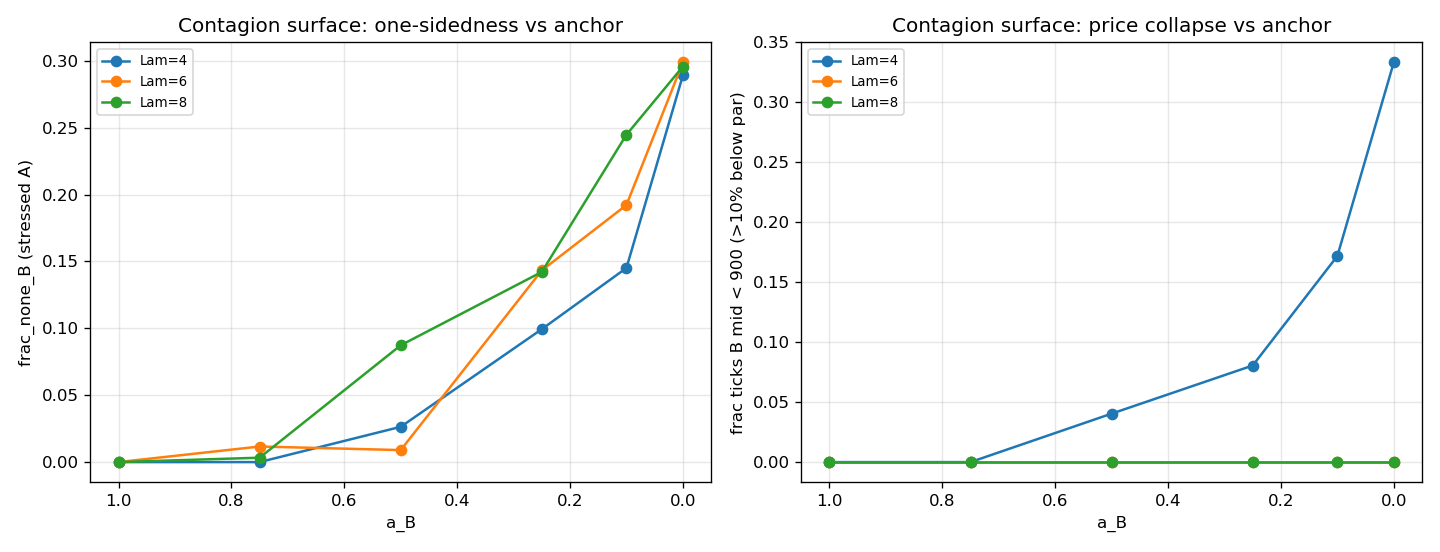}
\caption{Intrinsic stress rises as the anchor weakens. Under a stressed source, the receiver's
order parameter $\fnone^{B}$ climbs monotonically from a hard zero at full anchor ($a=1$) to
$\approx 0.30$ at zero anchor ($a=0$), and the leverage spiral's margin-breach rate rises from zero
to $\approx 1.8$ per tick. The dial reproduces the predicted instability, confirming the anchor as
the structural stabiliser.}
\label{fig:surface}
\end{figure}

The hierarchy establishes that the anchored receiver is resilient. To prove that the anchor is the
\emph{cause}, we remove it with the dial of Section~\ref{sec:anchordial}, sweeping the receiver's
anchor strength $a$ from $1$ to $0$ under the full coupling stack. The restoring force disappears
smoothly. The direct-shock test tells the story most sharply (Figure~\ref{fig:selfsustain}): at
$a=1$ the post-shock price reverts to the fundamental exactly and one-sidedness collapses back to
$\approx 0.04$, but as $a\to 0$ the post-shock one-sidedness rises roughly eightfold and stays
elevated after the shock relents--the dislocation no longer decays because nothing pulls the price
back. Correspondingly the receiver's intrinsic stress rises monotonically as the anchor weakens
(Figure~\ref{fig:surface}): under a stressed A, $\fnone^{B}$ climbs from a hard $0.000$ at $a=1$ to
$\approx 0.30$ at $a=0$, and the margin spiral now genuinely fires, its breach rate climbing from $0$
to $\approx 1.8$ per tick. Removing the anchor removes the resilience. This is the causal
confirmation that the fundamental anchor--and not some incidental feature--is the stabilising
restoring force.

\subsection{Intrinsic stability and contagion are separable}
\label{sec:separable}

\begin{figure}[t]
\centering
\includegraphics[width=0.82\textwidth]{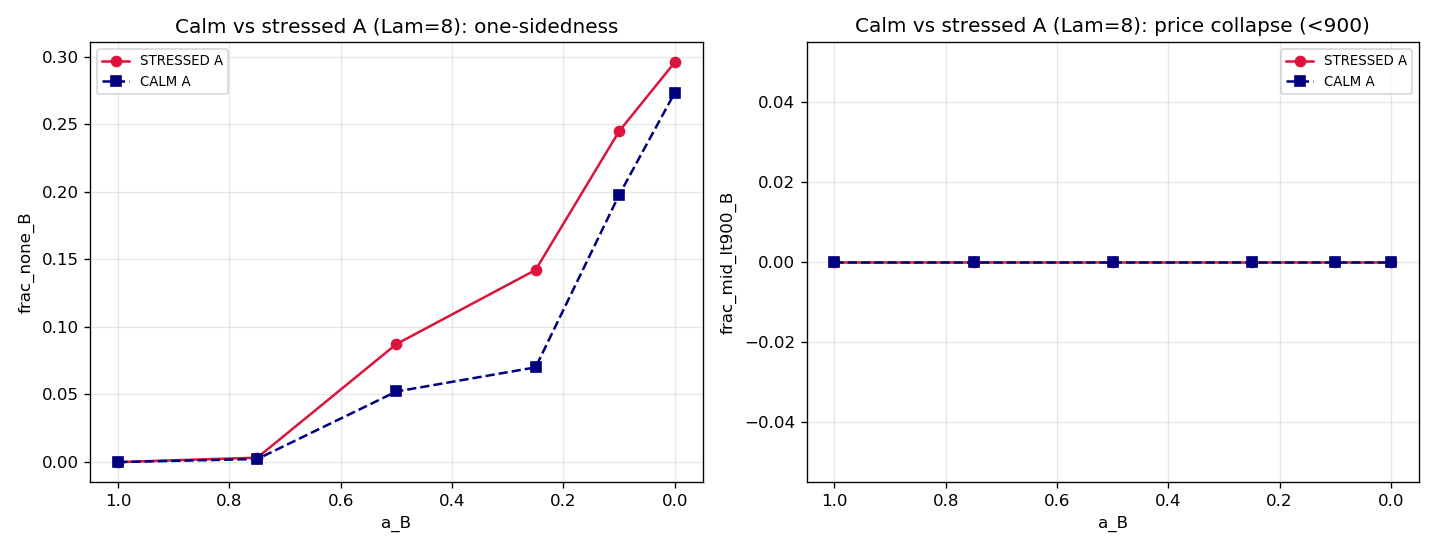}
\caption{Intrinsic fragility, not contagion. At every anchor strength the receiver's stress under a
stressed source (solid) and under a calm source (dashed) lie on top of each other; the increment is
statistically zero ($z$ from $-0.48$ to $-0.09$), with the calm-source curve if anything marginally
higher. Weakening the anchor makes the receiver unstable on its own, but the source market's stress
neither drives nor is required for that instability.}
\label{fig:calmvstressed}
\end{figure}

Weakening the anchor makes the receiver intrinsically fragile. It does not make the receiver catch
its neighbour's stress. This is the sharpest result of the paper, and it comes from the decisive
control run at every anchor strength. Even with the receiver unanchored and collapsing, a stressed A
drives no more instability than a calm A: at $a=0.25$, $0.10$, $0.00$ the stressed-minus-calm
increment in $\fnone^{B}$ is $-0.021$, $-0.017$, $-0.006$ with $z = -0.48$, $-0.43$, $-0.09$--the
calm-source curve lies, if anything, marginally above the stressed-source curve, and the two lie on
top of each other (Figure~\ref{fig:calmvstressed}). There is no window of anchor strength in which
the receiver is stable under a calm A but ignites under a stressed A, which is the signature genuine
contagion would leave. The receiver either refills (anchored) or destabilises on its own
(unanchored); it never ignites \emph{because} the source is stressed.

It is important to distinguish this from a related quantity that is easy to misread. At $a=0$ the
self-reinforcing spiral amplifies the receiver's stress well beyond a linear drain of the same
volume: the real spiral gives $\fnone^{B}=0.303$ against $0.102$ for a flow-matched synthetic drain,
an increment of $0.201$ at $z=3.78$. This is a genuine and reportable property of the mechanism ---
but it is a comparison of spiral against no-spiral \emph{within the receiver}, an \emph{intrinsic}
self-amplification of the unanchored market, present equally whether the source is stressed or calm.
It is not cross-market contagion. The adjudicator of transmission is the stressed-versus-calm control
above, and that is null. Reading the $z=3.78$ intrinsic amplification as contagion would be a
misattribution of cause. The two ingredient controls confirm the separation: at $a=0$ the withdrawal
channel is itself A-stress-independent (stressed $0.110 \approx$ unstressed $0.126$), and the spiral
fires the same with or without the cross-market priming (spiral-only $0.323 \approx$ full stack
$0.303$).

\section{Discussion}
\label{sec:discussion}

The picture that emerges is coherent. In a minimal order-book market whose liquidity is posted
around a fundamental value, the fundamental anchor is a restoring force that sets the market's
\emph{intrinsic} stability: it reverts a shocked price to the fundamental and refills the book faster
than a direct shock, a mechanical drain, a withdrawal-thinned book, or a leverage spiral can empty
it. We confirm this causal role directly, by dialling the anchor away and watching the intrinsic
stability vanish. Separately, and independently of the anchor, we find no cross-market contagion:
across the whole channel hierarchy and at every anchor strength, the receiver's stress does not
depend on whether the source is genuinely stressed. Intrinsic stability and cross-market contagion
are cleanly separable--the anchor governs the former, and the latter never appears at all.

\paragraph{Why the anchor wins.}
Every transmission channel we test acts, in the end, on the receiver's book: it removes liquidity
(the withdrawal channels) or consumes it (the order-flow and fire-sale channels). The anchor
replaces liquidity, and it does so at a fixed reference. A one-sided book is repaired by fresh
quotes at the fundamental; a dislocated price is pulled back to the fundamental; a forced seller can
always exit near par, so leverage losses stay bounded and no runaway can build. The receiver can be
stressed transiently--a violent shock does empty one side of its book--but the transient
decays because the restoring force is always on. This is why the strongest channel we build, a
funding-linked population that synchronously withdraws and a leverage spiral that self-reinforces
under a direct shock, still leaves the receiver's realised order parameter untouched. The mechanism
is present; the market is anchored against it.

\paragraph{A liquidity crisis is a failure of anchoring, not of market making.}
The strongest transmission channel we build is a synchronised, funding-constrained population of
market makers that pulls depth ($-64\%$) and widens spreads ($+136\%$) in the calm receiver because
a different market is stressed. This is exactly the procyclical liquidity withdrawal that the
flash-crash literature identifies as the engine of self-reinforcing collapse
\citep{cftcsec2010findings, karvik2018deeds, paulin2018flashcrash}. The narrative originates in the
primary regulatory post-mortem of the 2010 Flash Crash \citep{cftcsec2010findings}, yet the
empirical record already complicates a purely market-making-side account:
\citet{kirilenko2017flashcrash} find that the high-frequency traders did not literally withdraw as
prices fell but kept trading, aggressively consuming liquidity and passing inventory, which
strengthens rather than weakens our reading that withdrawal is necessary-looking but not sufficient.
In our fundamentally anchored market it thins the
receiving book but never ignites a crisis in it: the receiver's order parameter stays negligible and
independent of the source's stress. The self-sustaining collapse appears only when the fundamental
anchor itself is removed, at which point a leverage-driven fire-sale self-sustains without any
market-maker withdrawal being needed. This refines the procyclical-withdrawal narrative rather than
contradicting it: in a market with a working fundamental anchor, market-maker liquidity withdrawal
is necessary-looking but not sufficient to produce a crisis; what destabilises the book is the loss
of fundamental anchoring, and a genuine demand-side leverage spiral bites only once that anchor is
gone. In this vehicle, then, a liquidity crisis is a failure of fundamental anchoring, not of market
making. The closest empirical precedent for reading a crash as the failure of a \emph{link} rather
than of the dealers is \citet{menkveld2019flashcrash}, who attribute the Flash Crash to a breakdown
of cross-market arbitrage linkage rather than to a failure of market makers; our thesis is positioned
against theirs, not presented as unprecedented, and differs in kind--their broken link is inter-venue
arbitrage on a single asset in a single event, whereas ours is fundamental-value anchoring of resting
liquidity, dialled causally in a model. The scope is limited: the instability we produce is
one-sidedness of the book--
liquidity stress rather than a directional crash--and this is a minimal vehicle; the
destabilising ingredient we isolate is a demand-side leverage spiral acting on an unanchored book.

\paragraph{A separation the contagion literature can use.}
Our anchor-dial experiment makes explicit a distinction that is often conflated: a market that
\emph{can} collapse once you remove what was holding it up is not the same as a market that
\emph{catches} its neighbour's collapse. Removing the anchor produces the first--an intrinsically
fragile receiver that destabilises on its own, even under a completely calm neighbour--without
producing the second. Genuine contagion requires a channel that carries the source's stress into the
receiver, and none of ours does. The lesson for richer models is that a demonstrated withdrawal
mechanism, however large and crisis-like, is necessary but not sufficient evidence of contagion; the
adjudicating test is whether the receiver's stress depends on the source being genuinely stressed
rather than merely on the flow the coupling generates. Throughout, that test is null.

\paragraph{Scope and limitations.}
The positive claim is scoped narrowly. It is a claim about a stabilising restoring force
in this minimal zero-intelligence-plus-herding order-book vehicle, evidenced by the specific channels
we test; we do not claim universality beyond the vehicle. Several limitations bound the reading.
First, the demonstration that contagion is absent is inseparable from the stabiliser: to obtain a
receiver that could in principle catch a neighbour's stress, one must weaken the anchor, and the
weakened receiver is then intrinsically unstable under a calm neighbour--so the model cannot
exhibit a market that is stable alone yet infected by its neighbour. Whether that regime exists at
all requires a different coupling (for instance shared collateral marked to the receiver's own
falling price, or agents leveraged across both books at once) that carries the source's price into
the receiver's balance sheet, which we have not built. Second, the instability that appears at weak
anchor is genuine liquidity stress--one-sidedness of the book--but it is not a one-directional
crash: with the restoring force removed the price random-walks, drifting up as often as down, so the
order parameter of interest is the one-sided-book fraction rather than a downward price collapse.
Third, the demand-side fire-sale channel is limited by bounded market-maker inventory, which is why
the leveraged-holder population is required to build a first-order demand shock at all. Fourth, we
sweep one stressed source corner, one sub-critical receiver, and one maker/funding/holder
calibration; the calibration that keeps the calm receiver stable is the same one that bounds the
inventory a fire-sale can dump, a genuine tension we report rather than tune away. Finally, near the
receiver's own onset the order parameter is a rare-excursion, right-skewed quantity, so the small
point estimates in the withdrawal and fire-sale channels sit inside a seed-noise floor; the
anchor-sweep signals, by contrast, are large and clean, which is why the causal confirmation is the
firm part of the paper.

\section{Conclusion}
\label{sec:conclusion}

We have identified, quantified, and causally confirmed an intrinsic stabilising mechanism in an
order-book market: fundamental-value anchoring of liquidity provision acts as a restoring force that
reverts a shocked price to the fundamental and refills the book with two-sided depth after a shock.
Dialling the anchor down removes the restoring force, the receiver loses its mean-reversion, and a
leverage-driven fire-sale self-sustains--the direct causal confirmation that the anchor is the
stabiliser. Separately, we coupled two anchored markets and attempted to transmit a herding-driven
liquidity crisis from one into a calm neighbour through a hierarchy of progressively stronger
channels--a cross-market herding signal, arbitrage order flow, a risk-linked market maker, a
funding-constrained maker population that synchronously withdraws, a forced fire-sale, and a
leveraged-holder margin spiral. Although the transmission mechanism grew genuinely crisis-like across
the hierarchy, the receiver's stress never depended on whether the source was truly stressed, at
every anchor strength: no cross-market contagion appears.

The study draws a clean line between two properties that are easily conflated. The anchor governs a
market's \emph{intrinsic} stability: weaken it and the market becomes fragile on its own, even beside
a perfectly calm neighbour. Cross-market \emph{contagion} is a separate matter, requiring a channel
that carries the source's stress into the receiver, and no such transmission appears at any anchor
strength--the source market's stress is neither necessary nor sufficient for the receiver's
instability. In particular, market-maker liquidity withdrawal--even a synchronised,
funding-constrained population that pulls depth and widens spreads--thins the receiving book but
never ignites a crisis in it; the self-sustaining collapse appears only once the fundamental anchor
is removed. A liquidity crisis in this vehicle is a failure of fundamental anchoring, not of market
making. For richer models this suggests a concrete design principle and a concrete test: a restoring
force anchored to fundamentals is a stabiliser worth preserving, and a demonstrated
liquidity-withdrawal mechanism should be adjudicated as contagion only by whether the receiver's
stress tracks the source's stress rather than the coupling's flow. The natural next step is to build
the coupling the present vehicle deliberately lacks--one that marks a shared balance sheet to the
receiver's own price--and ask whether genuine transmission can be made to survive the same controls.

\section*{Code and data availability}
The model, the simulation and analysis code, and the data reproducing every figure and number in
this paper are available at \coderepo\ \citep{novotny2026sharedflowcode}.

\bibliographystyle{unsrtnat}
\bibliography{references}

\end{document}